# A Streaming Accelerator for Deep Convolutional Neural Networks with Image and Feature Decomposition for Resource-limited System Applications

Yuan Du [1,2*], Li Du [1,2], Yilei Li [1], Junjie Su [2], Mau-Chung Frank Chang [1,3]

[1] High Speed Electronics Lab (HSEL), University of California, Los Angeles, CA, USA, 90095
[2] Kneron Inc., San Diego, CA, USA, 92121
[3] National Chiao Tung University, Hsinchu, Taiwan, 30010
*yuandu@ucla.edu

**Abstract:** Deep convolutional neural networks (CNN) are widely used in modern artificial intelligence (AI) and smart vision systems but also limited by computation latency, throughput, and energy efficiency on a resource-limited scenario, such as mobile devices, internet of things (IoT), unmanned aerial vehicles (UAV), and so on. A hardware streaming architecture is proposed to accelerate convolution and pooling computations for state-of-the-art deep CNNs. It is optimized for energy efficiency by maximizing local data reuse to reduce off-chip DRAM data access. In addition, image and feature decomposition techniques are introduced to optimize memory access pattern for an arbitrary size of image and number of features within limited on-chip SRAM capacity. A prototype accelerator was implemented in TSMC 65 nm CMOS technology with 2.3 mm x 0.8 mm core area, which achieves 144 GOPS peak throughput and 0.8 TOPS/W peak energy efficiency.

## 1. Introduction

Deep convolutional neural networks (CNN) have shown significant performance and become ubiquitous in machine learning and artificial intelligence (AI) related applications, such as computer vision, speech recognition and smart security object recognition [1, 2]. There are plenty of applications and specific-domain hardware developed in high-performance server and datacentre field [5-7, 14-20]. However, this hasn't resulted in the wide use of deep CNNs on resource-limited platforms, such as mobile devices, internet of things (IoT), unmanned aerial vehicles (UAV), and so on. The reason is state-of-the-art CNNs requires more than tens of megabytes of parameters storage and billions of arithmetic operations in a single inference pass. Massive data movements between on-chip SRAM and off-chip DRAM and heavy computation resources are out of the cost and power budget of many resource-limited applications [8-14].

There are two main tasks needs to accomplished before modern deep CNNs could run in resource-limited platforms: (1) deeply compressing redundancy by pruning and quantization to simplify and reduce the size of networks [3]; (2) building moderate performance and high energy efficient hardware accelerator [8]. In this paper, we will focus on the latter one, and a streaming-based hardware architecture to accelerate the convolution layer and pooling layer in CNNs is proposed. It executes CNN computation not only to achieve high parallelism but also to optimize memory access pattern by maximizing local data reuse within limited bandwidth provided to off-chip DRAM, eventually to achieve high energy efficiency for the overall system. Also, image, feature and kernel decompositions make the proposed accelerator high-reconfigurable for different structures of networks.

## 2. CNN Basics and Layer Definition

A CNN (e.g., AlexNet, ResNet-18, VGG-16, etc.) usually comprises four typical layers: convolution layer (CONV), normalization layer (NORM), pooling layer (POOL), and fully-connected layer (FC). Modern CNNs can achieve better performance by utilizing a deep hierarchy of layers, where CONV layer computation accounts for more than 90% of the CNN operations and dominates runtime. Although FC layer use most of the filter weights, yet these weights are largely compressible to 1-5% of their original size [4]. Consequently, this paper will focus on CONV and POOL acceleration.

CONV applies convolution function to convert input layer's image/feature map to the next layer's feature map. Since each input layer can have multiple features, the convolution is 4-D, as shown in Fig. 1. Each filter or feature map is a 3-D structure with multiple 2-D planes, and a group of 3-D feature maps are multiplied by a group of 3-D filters and added by a 1-D bias vector. The computation of CONV layer is defined as Equation (1).

$$O[o][m][x][y] = B[o] + \sum_{k=1}^{M}\sum_{i=1}^{K}\sum_{j=1}^{K} I[o][k][\alpha x + i][\alpha y + j] \times W[m][k][i][j]$$
$$1 \leq o \leq N, 1 \leq m \leq M, 1 \leq x,y \leq S_o \quad (1)$$

$O$, $B$, $I$, and $W$ are the output features, biases, input features, and filters, respectively.

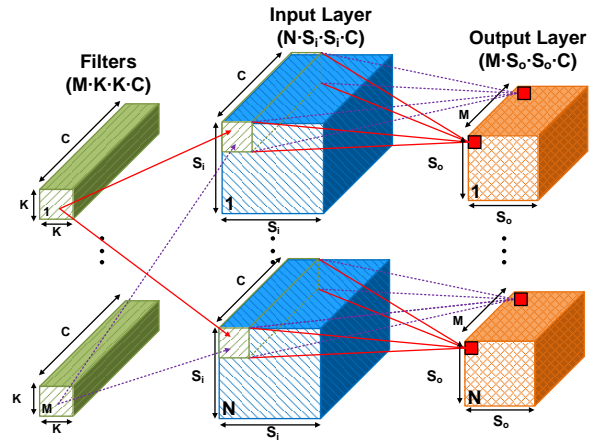

**Fig. 1.** *Concept of computation of CONV layer*

## 3. Streaming Architecture

To minimize data movement and to utilize maximal available on-chip SRAM bandwidth, a streaming architecture is proposed, as shown in Fig. 2. SRAM width is set to 16 Byte, corresponding to stream 8 pixels per cycle.

As Fig. 2(a) shows the proposed single channel column buffer with 2 x N row buffer (N is the depth of SRAM), which solves the input data boundary issue and make input data bandwidth the same with CUA output bandwidth. As a result, the convolution computation process is continuous and stream-like. There is no need to pause or wait for the incomplete convolution calculation. In Fig. 2(b), after the first eight rows, every cycle has eight groups' valid convolution results with the help of 2xN pixel size ROW BUF.

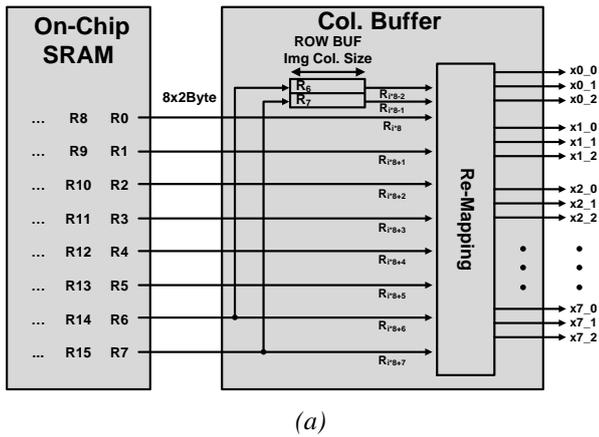

(a)

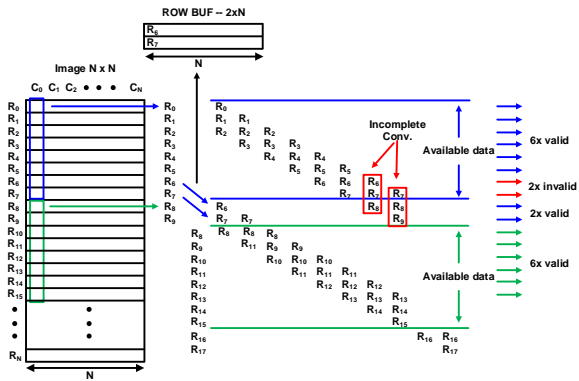

(b)

***Fig. 2.*** *An example of streaming architecture (a). Column buffer architecture; (b). Streaming data flow*

## 4. System and Building Blocks Design

### 4.1. Top-level Architecture.

The overall system-level architecture is illustrated in Fig. 3. It includes a 128 Kbyte Single Port SRAM as the Buffer Bank to store intermediate data, exchange data with the DRAM. A column buffer module (COL BUFFER) is implemented to remap the SRAM output to the convolution engine array input. CU Engine array is composed of sixteen 3x3 convolutional unit to enable highly parallel convolutional computation. The filter coefficient is pre-fetched directly using a pre-fetch controller inside the engine to periodically update the weight and bias value.

Finally, an accumulation buffer with partial summation and max pooling function included accumulating the partial convolution results coming from the CU engine and pool the final convolution output if necessary. The control of this accelerator is through 16-bit AXI bus; the command decoder is integrated inside the accelerator. The commands for the processed CNN net is pre-stored in the DRAM already and will be automatically loaded to a 128-depth command FIFO during power up.

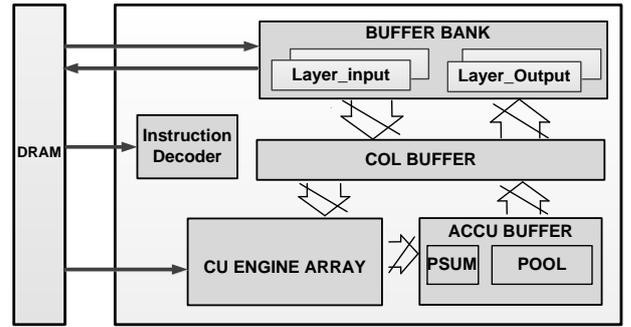

***Fig. 3.*** *System-level architecture of CNN accelerator*

### 4.2. CU Engine Array

The CU engine array includes nine processing engines (PE) and an adder to combine the output, as shown in Fig. 4. The PE acts as a multiplication function between the input data and the filter coefficient and meanwhile pass the input data to the next stage's PE's input through a D flip-flop. The multiplication function can be turned on/off based on the EN_Ctrl signal to save the computation power when convolution stride size is larger than one. As described in Section 4.1, the accelerator uses nine multipliers to form a CU and sixteen CUs to compose a CU engine.

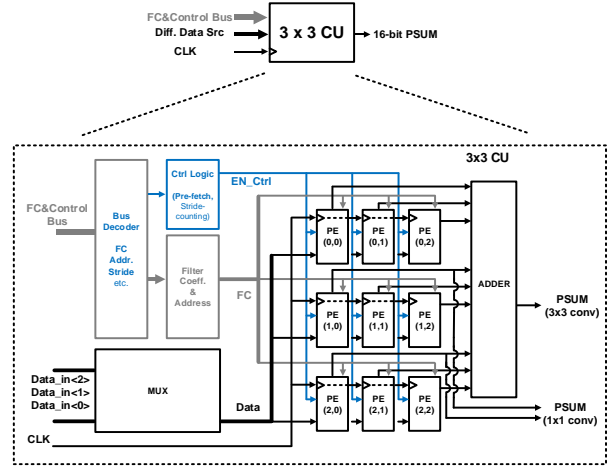

***Fig. 4.*** *Convolution engine implementation*

In the 3x3 convolution, the multiplied result will send to the adder in the CU to perform the summation and deliver the summed result to the final output. Filter weights will be fetched from the DRAM through the DMA controller and pre-stored in the CU through a global bus. When one channel is scanned, a synchronized filter updated request signal will be sent to the CU to update the filter weights at the PE's input for the upcoming channel.

### 4.3. Reconfigurable Streaming-based Pooling Block

Fig. 5 shows an overview architecture of the pooling module and its connection to the scratchpad. The scratchpad stored row data from one output feature in parallel. The row data share one column address and can be accessed simultaneously. Because of the difference of stride size in the convolution, data stored in the scratchpad may not be all validated. For example, when the stride is equal to 2, only R0, R2, R4, R6 store the validate data. In addition, the pool window's kernel size can also be configured to be 2 or 3.

To accommodate different convolution strides and pool-size cases, a multiplexer is put in front of the max pooling module to select the validated input data to the corresponding max-pool units. The max-pool unit is implemented with a four-input comparator and a feedback register to store the intermediate comparator output result. In addition, an internal buffer is embedded in the max pooling module. This is to buffer the intermediate results if some of the data inside the pooling window are not ready.

When a pooling begins, the comparator first takes three input data coming from nearby rows (two data in 2x2 case) and output the maximum value among the input data. This temporary maximum value will be fed back to the comparator's input and regarded as one additional input to compare with the next clock cycle's input data. This procedure will be duplicated till the whole pooling window's input data is scanned. After that the output enabling signal will be validated and output the maximum value in the pooling window.

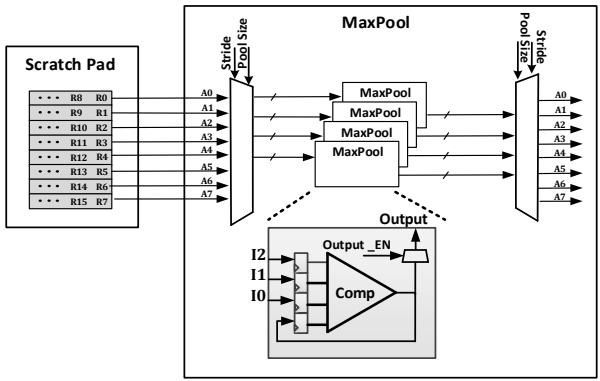

*Fig. 5.* Overall architecture of the pooling module, Ri represents row i's data. The pooled output will be fed back to the scratchpad.

## 5. Image and Feature Map Decomposition

For source-limited embedded systems, image/feature decomposition technique is significantly necessary to support high-resolution images, large size and a large number of feature maps, simultaneously maintaining peak throughput, which is limited by the amount of CU engine array or off-chip DRAM bandwidth. Table 1 summaries the computation and storage requirements for AlexNet [1, 21-23].

**Table 1** AlexNet operations and storage summary

| Layer # | Input Layer Size | Output Layer Size | Num. Ops | Input Mem. | Output Mem. | Total Mem. |
|---|---|---|---|---|---|---|
| 1 | 227x227x3 | 55x55x96 | 211M | 309KB | 581KB | 890KB |
| 2 | 27x27x96 | 27x27x256 | 448M | 140KB | 373KB | 513KB |
| 3 | 13x13x256 | 13x13x384 | 299M | 87KB | 130KB | 216KB |
| 4 | 13x13x384 | 13x13x384 | 224M | 130KB | 130KB | 260KB |
| 5 | 13x13x384 | 13x13x256 | 150M | 130KB | 87KB | 216KB |
| Total | | | 1.3G | 0.8MB | 1.3MB | 2.1MB |

Taking the 1st layer of AlexNet, shown in Fig. 6, the input image is decomposed into nine parts, so the input layer on-chip SRAM size is reduced to 34 KB. Accordingly, the output layer on-chip SRAM size reduced to 33 KB with the help of both image and feature decomposition by 9 and by 2, respectively. The decomposition enables the source-limited hardware to accelerate arbitrary size and feature number of different CNN models at the cost of slower computation.

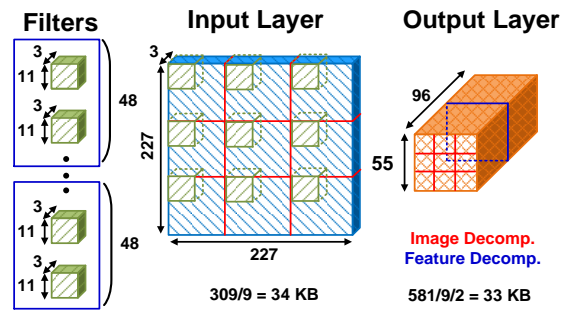

*Fig. 6.* Image, feature decomposition concept of the 1st layer of AlexNet

## 6. Accelerator Implementation

The accelerator is implemented in TSMC 65 nm CMOS GP standard VT technology. To evaluate the area, power and critical path, we developed the register-transfer level (RTL) design in Verilog, then it is synthesized using Synopsys Design Compiler (DC) Version G-12.06-SP1. We placed and routed the accelerator using Cadence Innovus Version 16.10. We used ARM Artisan Physical IP to generate SRAM RTL model and GDS layout.

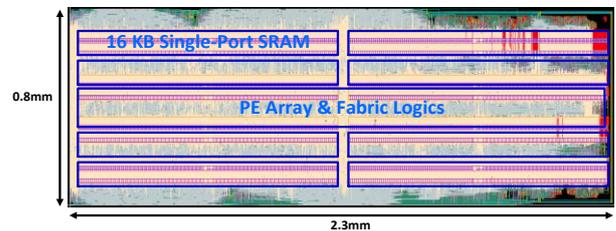

*Fig. 7.* Layout view and area breakdown of the proposed accelerator in TSMC 65nm CMOS technology

The layout is shown in Fig. 7. 57%, 35% and 8% of the area are taken by SRAM buffer bank, CU engineer array, and column buffer, respectively. The core area is 2.3 mm x 0.8 mm, achieves a peak throughput of 144 GOP/s at a 500 MHz clock and a peak energy efficiency 0.8 TOPS/W at a 20 MHz clock. The detailed performance is summarized in Table 2.

**Table 2**: Performance summary

| | |
|---|---|
| Technology | 65nm CMOS |
| Supply Voltage | 0.6~1 V |
| Clock Rate | 20MHz ~ 500MHz |
| Power | 7 mW @ 20 MHz & 0.6 V<br>425 mW @ 500 MHz & 1.0 V |
| Area | 2.3 mm x 0.8 mm |
| Gate Count | 0.3 million |
| Number of CU Eng. | 16 (9 PEs per CU Eng.) |
| On-chip Single Port SRAM | 128 KB |
| Precision | 16-bit fixed point |
| Throughput | 144 GOPS @ 500 MHz<br>5.8 GOPS @ 20 MHz |
| Energy Efficiency | 0.3 TOPS/W @ 500MHz<br>0.8 TOPS/W @ 20MHz |

The accelerator RTL design is also verified on Xilinx ZCU102 valuation platform. Filter weights will be fetched from the DRAM through the DMA controller and pre-stored in the CU through a global bus. The application processor (AP) integrated into the FPGA is used to control the accelerator and initiate the computation. Through using the DMA controller inside the FPGA, the accelerator can successfully access the data and the weights stored in the DRAM. The demonstration setup is shown in Fig. 8.

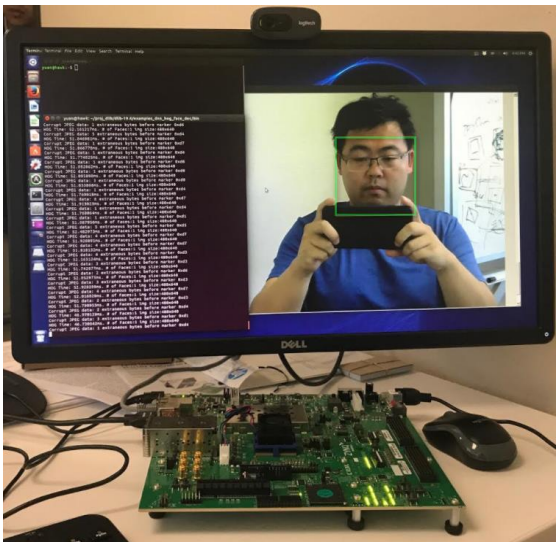

***Fig. 8.*** *Face detection demonstration on the Xilinx ZCU102 FPGA*

## 7. Conclusion

In this paper, a streaming-based accelerator for deep CNNs with image and feature decomposition technique is proposed to accelerate convolutional neural network computation. It also supports arbitrary sizes and feature numbers to fit source-limited embedded platform. In addition, pooling function is also supported in this accelerator through integrating separate pooling module and proper configuration of the convolution engine. It is able to support most popular CNNs, and achieve 0.3 TOPS peak throughput and 0.8 TOPS/W peak energy efficiency in TSMC 65nm CMOS technology with a core size of 1.84 mm$^2$.